\documentclass[conference,a4paper]{IEEEtran}
\usepackage{amssymb,amsmath}
\usepackage{cite}
\usepackage{psfrag}
\usepackage{url}
\usepackage[latin1]{inputenc}
\usepackage[absolute,overlay]{textpos}

\newcommand{\argmin}[1]{{\operatorname{arg}\,\min_{#1}}\,}
\usepackage{pgf}
\usepackage[latin1]{inputenc}
\usepackage[export]{adjustbox}

\usepackage{amsthm}

\begin{document}

\title{Enabling URLLC for Low-Cost IoT Devices via Diversity Combining Schemes}
\author{
	\IEEEauthorblockN{Onel L. Alcaraz López, Nurul Huda Mahmood, Hirley Alves}
	\IEEEauthorblockA{Centre for Wireless 	Communications (CWC), Oulu, Finland\\}
				\{onel.alcarazlopez,nurulhuda.mahmood,hirley.alves\}@oulu.fi}

\maketitle

\begin{abstract}
	Supporting Ultra-Reliable Low-Latency Communication (URLLC) in the Internet of Things (IoT) era is challenging 
	due to stringent constraints on latency and reliability combined with the simple circuitry of IoT nodes.
	Diversity is usually required for sustaining the reliability levels of URLLC, 
	but there is an additional delay associated to auxiliary procedures to be considered, specially when communication includes low-cost IoT devices. 
	Herein, we analyze Selection Combining (SC) and Switch and Stay Combining (SSC) diversity schemes as plausible solutions for enabling ultra-reliable low-latency downlink communications to low-cost IoT devices. We demonstrate the necessity of considering the time spent in auxiliary procedures, which has not been traditionally taken into account, while we show its impact on the reliability performance. 
	We show there is an optimum number of receive antennas, which suggests that under certain conditions it might be required to turn off some of
	them, specially under the SC operation.
	We highlight the superiority of SSC with respect to SC as long the associated Signal-to-Noise Ratio threshold is properly selected. We propose using a fixed threshold relying only on long-term channel fading statistics, which leads to near-optimum results.
%
\end{abstract}

\section{Introduction}\label{intro}
The Internet of Things (IoT) paradigm enables 
several use-cases 
by combining the physical sensing of IoT devices with data analysis. Aided by ubiquitous wireless connectivity, declining costs, and the emergence of cloud platforms, our world is moving towards the IoT era \cite{Dhillon.2017}. The fifth-generation New Radio standard (5G NR) targets for the first time service classes related to Machine-Type Communication (MTC), which constitute the IoT's core. 
Specifically, MTC use cases related to ultra-reliable low-latency communication (URLLC) are very challenging as they are characterized by extremely stringent requirements \cite{3GPP.2017}, {e.g.} \cite{Schulz.2017}: factory automation, with maximum latency around $0.25-10$ms and maximum error probability of $10^{-9}$; smart grids ($3-20$ms, $10^{-6}$), professional audio ($2$ms, $10^{-6}$), etc.
Notice that guaranteeing simultaneously low-latency and high-reliability is what makes URLLC design very complicated \cite{Ji.2018}.

In general, diversity is required to sustain the desired reliability levels in URLLC \cite{Popovski.2018}. Among the most promising diversity techniques are those relying on the spatial domain by  incorporating multiple antennas at the transmitter and/or receiver. Maximum Ratio Combining (MRC) and Selection Combining (SC) have been by far the most popular receive diversity schemes in the last two decades. 
However, when the receiver is a low-cost IoT device, e.g. a radio frequency (RF) energy harvesting device \cite{Lopez.2017} or even most of sensors, they may not be appropriate. The reasons are that the multiple RF chains, which would be required for implementing for instance MRC and certain SC variants, are often unaffordable because of the required extremely \textit{i)} small form-factor and/or \textit{ii)} low power consumption.
Thus, switching-like combining schemes such as Switch and Stay Combining (SSC)\footnote{Notice that SSC was originally conceived for a two-antenna node with the possibility of a unique switching. In this work, we refer to a more general implementation, which may also be found in the literature as Switch-and-Examine Combining (SEC) \cite{Yang.2003} or even SEC with post-examining selection \cite{Yang.2006}.} or single RF chain SC, are necessary 
to still benefit from multiple receive antennas to provide low-power URLLC in (and beyond) 5G wireless systems.

Literature on the performance of both SC and SSC either in terms of reliability, throughput, energy efficiency or switching rates, is large, e.g. \cite{Goldsmith.2005,Yang.2003,Yang.2006,Beaulieu.2008,Avendi.2013,Lopez2.2018,Lopez.2019}. 
%
However, to the best of our knowledge all works assume sufficiently long messages such that the time consumed during the required auxiliary procedures, e.g. synchronization, Channel State Information (CSI) acquisition, signal processing, switching\footnote{Readers can refer to \cite{Soujeri.2016}, where authors evaluate the detrimental effect of the antenna switching time for spatial modulation systems.}, etc, can be ignored. Obviously, this is no longer the case in URLLC IoT use cases for which messages are short and must be delivered with ultra-reliability within strict time deadlines. Notice that the time spent in these auxiliary procedures increases with the number of receive antennas $M$, therefore, it does not hold anymore that by increasing $M$, the reliability performance improves unbounded, as occurs in delay-tolerant systems. 

This paper tackles above issues by considering the auxiliary procedures' time when analyzing the performance of downlink communications to a multi-antenna IoT device equipped with a single RF chain. The main contributions are: \textit{i)} we provide an analytical framework for evaluating the reliability performance of switching-like combining schemes under stringent latency constraints, \textit{ii)} for SSC we propose using a fixed Signal-to-Noise Ratio (SNR) threshold  that depends on the channel fading statistics, which allows reaching near-optimum results, \textit{iii)} we show numerically the existence of an optimum $M$, which suggests that under certain conditions it might be required to turn off some antennas.  Our results evidence how misleading are the traditional asymptotic approaches where auxiliary procedures are neglected, 
and show that SSC can always outperform SC, specially under very stringent latency constraints.

Next, Section~\ref{system} presents the system model and assumptions. Section \ref{SC} and \ref{SSC} discuss the reliability performance of SC and SSC diversity combining schemes, respectively, under stringent delay constraints. Finally, Section~\ref{results} presents numerical results and Section~\ref{conclusions} concludes the paper.
%
%
\section{System Model}\label{system}
Consider a downlink communication under which an urgent message consisting of $k$ information bits must be reliably delivered within a latency window of $u$ channel uses to a low-cost IoT device equipped with $M\ge 2$ antennas. The bandwidth is normalized such that the latency would be $uT_c$ in time units by assuming the symbol time duration is $T_c$.

 Although multiple antennas are available, we consider a unique RF chain due to simple hardware, small form-factor and low-energy profile requirements, which are typical of IoT setups. Therefore, the RF chain must be switched between the antenna elements  to take advantage of spatial diversity, and consequently switching-like diversity combining schemes are required. However, the simple hardware,  and stringent time deadline specified by a relatively small $u$, make it necessary to consider the impact of the antenna switching and processing time in the analysis and scheme design. In that sense, let $p\ge 1$ be the number of channel uses spent when switching the RF chain from one antenna element to another, while $q\ge 1$ accounts for the channel uses required for measuring the SNR of the signal in the active antenna, which includes also the involved signal processing tasks and consequent decision making\footnote{While the time required for decoding the metadata signal at each antenna is included in $q$, the decoding time of the information signal, $u_d$ (in channel uses), can be considered straightforwardly by setting $u\leftarrow u-u_d$. In general, $u_d$ is influenced by the particular hardware platform and depends on the payload to be decoded and the decoding algorithm. However, for synchronization feasibility it is desirable to use a fixed $u_d$, thus, an upper-bound value would be the most appropriate for projecting the system. Readers can refer to \cite{Celebi.2019}, where $u_d$ is modeled and upper-bounds are provided.}. We assume perfect CSI at the receiver\footnote{Notice that estimating the CSI can be carried out during a portion of the $q$ channel uses. Although such CSI is prone to errors due to the hard latency constraint, including the effect of imperfect CSI would demand a more elaborated mathematical analysis that is out of the scope of this work.} and quasi-static fading such that channel remains constant during the transmission of a message. Finally, the antenna elements are sufficiently separated such that the fading is independently and identically distributed throughout them. The analysis under correlated fading channels is left for future work.

As communication takes place over a small number $n<u$ of channel uses, we consider the error probability at finite blocklength (FB) as the reliability performance metric. For an Additive White Gaussian Noise (AWGN) channel with SNR $\gamma$, the error probability is 
\cite{Polyanskiy.2010}
\begin{align}\label{e1}
\varepsilon(k,n,\gamma)\approx Q\left(\frac{C(\gamma)-k/n+\frac{1}{2n}\ln(n)}{\sqrt{V(\gamma)/n}}\right),
\end{align}
where $C=\log_2(1+\gamma)$ and $V=\big(1-\frac{1}{(\gamma+1)^2}\big)\log_2^2e$ are the capacity and dispersion of the channel, respectively, %
and $Q(x)=\int_{x}^{\infty}\frac{1}{\sqrt{2\pi}}e^{-t^2/2}\mathrm{d}t$.
For quasi-static fading channels we can departure from \eqref{e1} to write
\begin{align}\label{e2}
\varepsilon(k,n)=\mathbb{E}_\gamma \big[\varepsilon(k,n,\gamma)\big]
\end{align}
since such channel can be interpreted as conditionally Gaussian given the SNR. However, it has been shown in \cite{Mary.2016} (and corroborated in \cite{Lopez2.2018} for a diversity-combining setup) that the effect of the fading makes the impact of the FB to vanish when $k$ is not extremely small and/or there is not a strong Line of Sight (LOS) component. Regarding the latter notice that diversity becomes in fact more relevant as the LOS decreases.
Hence, in such cases the asymptotic outage probability is a good match, thus 
\begin{align}\label{e3}
\varepsilon(k,n)&\approx \mathbb{P}\big[\gamma<2^{k/n}-1\big]\nonumber\\
&=F_\gamma\big(2^{k/n}-1\big).
\end{align}
The intuition behind this result is that the dominant error event over quasi-static fading channels is that the channel is in a deep fade. Since the transmitted symbols experience all the same fading, it follows that coding is not helpful against deep fades in the quasi-static fading scenario, hence, the FB error probability and the outage probability are already close to each other for small blocklengths.

Next, we describe and analyze the SC and SSC combining schemes in the context of the URLLC scenario considered in this work.
\section{Selection Combining (SC)}\label{SC}
Under the SC scheme all antenna branches are scanned
and the combiner  outputs the signal on the antenna branch with the highest SNR. Then, the SNR at the antenna that finally becomes  active is $\gamma_\mathrm{sc}=\max(\gamma_1,\cdots,\gamma_M)$, where $\gamma_i$ is the SNR at the $i-$th antenna; while its Cumulative Density Function (CDF) and Probability Density Function (PDF) are
%
\begin{align}
F_{\gamma_\mathrm{sc}}(x)&=\mathbb{P}[\gamma_\mathrm{sc}<x]=
F_{\gamma}(x)^M,\label{F}\\
f_{\gamma_\mathrm{sc}}(x)&=\frac{d}{d x}F_{\gamma_\mathrm{sc}}(x)=
MF_{\gamma}(x)^{M-1}f_{\gamma}(x).\label{f}
\end{align}

Notice that for $M\ge 2$ the transmitter consumes $(p+q)M$ channel uses before transmitting the data message\footnote{The device needs $M-1$ switches to test all antennas and one more to go back to the best one. Note that even if the last one happens to be the best, the transmitter does not know about it and needs to assume that the $M-$th switching occurred.}. 
Therefore, the number of channel uses over which the communication takes place is 
\begin{align}\label{nsc}
n_{\mathrm{sc}}=u-(p+q)M,
\end{align}
while operation is only feasible if $u>(p+q)M$.
Now, by taking advantage of \eqref{e2}, \eqref{e3}, \eqref{F}, the average error probability under the SC operation can be written as
\begin{align}
\varepsilon_\mathrm{sc}&=\mathbb{E}_{\gamma_\mathrm{sc}} \big[\varepsilon(k,n_\mathrm{sc},\gamma)\big],\label{Fg0}\\ 
&\approx F_{\gamma_\mathrm{sc}}(2^{k/n_\mathrm{sc}}-1) = F_{\gamma}(2^{k/n_\mathrm{sc}}-1)^M,\label{Fg}
\end{align}
where $n_\mathrm{sc}$ is given in \eqref{nsc}.
%
\section{Switch and Stay Combining (SSC)}\label{SSC}
Under the SSC scheme, the antenna branches are scanned in sequential order until the SNR is above a given threshold $\gamma_0$. Once a branch is chosen, as long as the SNR on that branch remains above $\gamma_0$, the combiner outputs that signal; while when the SNR on the selected branch falls below such threshold, the combiner switches to another branch. For fixed transmit rate setups and considering infinite blocklength (IB) transmissions, it is necessary to set $\gamma_0=2^{k/u}-1$ for the best reliability performance because \textit{i)} the transmit rate on each transmission approximates accurately to $k/u$ since $u\gg p,q:u\approx n$, and \textit{ii)} overpassing such threshold guarantees an error-free communication. In fact, the reliability performance of both SC and SSC would match since only when the maximum SNR exceeds the threshold $2^{k/u}-1$, SSC finds one antenna branch with SNR above it. However, in FB regime none of above situations hold since $u$ and $n$ may differ significantly, and even when operating with SNR above $2^{k/n}-1$ there are still chances of errors, as captured by \eqref{e1}. 

Another important issue is: \textit{what if the SNR at all the antennas is below $\gamma_0$?} Under the asymptotic formulation with $\gamma_0=2^{k/u}-1$, such situation conduces to an outage, while under FB transmissions there are chances of succeeding even when a SNR is below the threshold (even the asymptotic threshold). Therefore, a last switching to the antenna that performs the best is advisable. 
Finally, SSC scheme requires a feedback channel to inform the transmitter when to start data transmission since such information depends on the number of antenna switches. Such feedback is carried out out-of-band and utilizing $d\ge 1$ channel uses over an error-free channel. Although last assumption may seem too ideal, one must notice that the feedback information may consist of one or few bits and by considering an appropriate value of $d$ the error probability of such feedback channel could be several orders below the actual information  error probability, thus, mimicking an error-free channel.

Now, the average error probability under the SSC operation can be written as
%
\begin{align}\label{essc}
\varepsilon_\mathrm{ssc}(\gamma_0)=&\underbrace{\big(1-F_{\gamma}(\gamma_0)\big)\sum_{i=1}^{M}\bigg(F_{\gamma}(\gamma_0)^{i-1} \varepsilon\big(k,n_{i-1}\big)\bigg)}_{T_1}+\nonumber\\
&\qquad\qquad\qquad\qquad+\underbrace{F_\gamma(\gamma_0)^M \varepsilon\big(k,n_M\big)}_{T_2},
\end{align}
where
\begin{align}
n_i&=\left\{ \begin{array}{ll}
u-\big(z_i+d_i\big), & i=0,\cdots, M-1 \\
n_\mathrm{sc}, &  i=M
\end{array}
\right.,\label{ni}\\
z_i&=(i+1)q+ip,\\
d_i&=\left\{ \begin{array}{lc}
d, &   d<u-z_i \\
\\ (M-i)p+(M-i-1)q, & d\ge u-z_i
\end{array}
\right.,\label{d}
\end{align}
for $i=0,\cdots,M$, while $n_\mathrm{sc}$ is given in \eqref{nsc}. Notice that $z_i$ is the number of channel uses consumed during $i$ antenna switches, while $d_i$ and $n_i$ represent the \textit{actual} feedback delay and the number of channel uses available for transmission, respectively, at that point. The notion of actual feedback delay is introduced to deal with those situations in which after  switching some times, $d$ is  already relatively large and sending a feedback message would consume all the remaining time resources (or even more). In such cases it is preferable waiting $M(p+q)-(i+1)q-ip$ channel uses, which matches the second line in \eqref{d}, after which the transmitter would send automatically the data message by inferring that all antennas were already scanned, as in the SC scheme. 
Back to \eqref{essc}, notice that the first term $T_1$ encompasses  all the error events when the SNR is above the threshold $\gamma_0$ in at least one antenna, while the second term $T_2$ covers those events in which a last switching is still required since the SNR was below $\gamma_0$ in all the antennas.
It is important to note that for computing the FB error terms inside the summation we need to calculate $\mathbb{E}_\gamma \big[\varepsilon(k,n,\gamma)\big]$ over $\gamma\in [\gamma_0,\infty)$ using PDF $f_{\gamma}(x)/(1-F_{\gamma}(\gamma_0))$; while the FB error within $T_2$ must be evaluated by integrating over the interval $[0,\gamma_0)$ and using the PDF $f_{\tilde{\gamma}}(x)=f_{\gamma_\mathrm{sc}}(x)/F_{\gamma_\mathrm{sc}}(\gamma_0)$ where $\tilde{\gamma}\triangleq\big\{\max(\gamma_1,\cdots,\gamma_M)\big|\{\gamma_1,\cdots,\gamma_M<\gamma_0\}\big\}$. 

To optimize the SSC performance, the threshold $\gamma_0$ must be properly set. While a small $\gamma_0$ avoids the chances of unnecessary switches, it may also cause staying on a poor-SNR antenna branch since most of the antennas were not scanned, raising an interesting trade-off. Hence, we can establish the following optimization problem
%
	\begin{align} \label{P1}
	\mathbf{P:}\qquad \argmin{\gamma_0}       \   
	\varepsilon_\mathrm{ssc}(\gamma_0).
	\end{align}
%
Notice that the optimized error probability of SSC is upper-bounded by SC's since by setting $\gamma_0=\infty$ both \eqref{essc} and \eqref{Fg0} match. Therefore, SSC will always perform superior than SC as long as the SNR threshold is properly selected.
Since we are interested in the ultra-reliability region for which $\varepsilon\ll 1$, it is obvious that $\gamma_0>2^{k/u}-1$, but beyond such fact solving $\mathbf{P}$ is extremely complicated due to the convoluted form of \eqref{essc}. Next, we provide two plausible approaches.
\subsection{Naive approach (NA)}
%
%
Our approach here exploits the fact that an $M-$th switching would not impact significantly the reliability performance if the SNR at all antennas is below such threshold since the error probability would be already above $1/2$. In that sense it seems wise setting $\gamma_0=2^{k/n_M}-1$.
Notice that
even when the SNR is closely above such $\gamma_0$ during the first switchings, the error probability would be much smaller than that after the $M-$th switching since the number of channel uses could be significantly greater than $n_M$. 
This approach aims at reducing the impact of the $M-$th switching on the overall error probability in \eqref{essc}. This is because $$F_\gamma(\gamma_0)^M\!\int\limits_{0}^{\gamma_0}\!\varepsilon (k,n_M,x)\frac{f_{\gamma_\mathrm{sc}}(x)}{F_{\gamma_\mathrm{sc}}(\gamma_0)}\mathrm{d}x\!=\!\!\int\limits_{0}^{\gamma_0}\!\varepsilon (k,n_M,x)f_{\gamma_\mathrm{sc}}(x)\mathrm{d}x$$ is an increasing function of $\gamma_0$, hence, a value greater than $2^{k/n_M}-1$ would increase the impact of $T_2$ in \eqref{essc}.

Finally, the method proposed here avoids using any information on the channel fading, hence the \textit{naive approach} term; and consequently could lead to far-from-optimal results in many cases.

\subsection{Proposed Fading-dependent approach (FA)}
Given $n$ transmit channel uses to communicate over any of the $M$ antennas, the average error probability is asymptotically lower-bounded by $F_{\gamma}(2^{k/n}-1)^M$.
Thus, it follows immediately that the SSC error probability is lower-bounded by $F_{\gamma}(2^{k/n_0}-1)^M$, where $n_0$ represents the number of channel uses available for transmission if the first antenna is selected for transmission, and it can be computed according to \eqref{ni}. 
Notice now that the instantaneous error probability is not bounded by such minimum average error probability.
According to this, and since the ultimate goal of an efficient SSC scheme lies in finding a high-SNR antenna branch in relative few switches, it is necessary targeting a greater error probability as threshold. Hence, instead of exponent $M$, we will use one order of magnitude lower exponent, e.g.  $F_{\gamma}(2^{k/n}-1)^{M-1}$ as the target error probability.

Another issue is that the number of available transmit channel uses decreases after each switching, while increasing the attainable lower-bound error probability. We avoid using different SNR thresholds at each switching\footnote{Designing an efficient SSC scheme using a different SNR threshold at each switching is a challenging task, though no performance gains are expected.} and consequently we adopt the $l-$factor generalized mean of the $n_i$s as 
%
\begin{align}
\tilde{n}=\Big(\frac{1}{M+1}\sum_{i=0}^{M}n_i^{l}\Big)^{1/l},\ \  l\in\mathbb{R},
\end{align}
as an equivalent $n$; while using it for calculating the error probability threshold when scanning the antenna branches. 
This approach is advantageous since by tuning the factor $l$ we allow $\tilde{n}$ to take values in the interval $[n_0,n_M]$ and the target error probability
\begin{align}
\xi=F_{\gamma}(2^{k/\tilde{n}}-1)^{M-1}
\end{align}
in the interval $\big[F_{\gamma}(2^{k/n_0}-1)^{M-1}, F_{\gamma}(2^{k/n_M}-1)^{M-1}\big]$. 
Obviously,  it is presumably preferable to operate under SC if $\xi\ge \varepsilon_\mathrm{sc}$, thus, $\gamma_0=\infty$.
In such case, SC and SSC are equivalent. 
Otherwise,  the problem translates to finding the SNR for which the error probability becomes $\xi$ according to \eqref{e1}. Notice that inverting \eqref{e1} for $\gamma$ is analytically intractable, thus we resort to \cite{Lopez.2018}
 \begin{align}
 \gamma_0^{(t)}=2^{\tfrac{k}{\tilde{n}}-\tfrac{1}{2\tilde{n}}\ln(\tilde{n})+\sqrt{\tfrac{1}{\tilde{n}}V\big(\gamma_0^{(t-1)}\big)}Q^{-1}(\xi)}-1,
 \end{align}
 and use $\gamma_0^{(0)}=\infty$. In fact, the number of required iterations $t$  is very small in most of the scenarios \cite{Lopez.2018}. Finally, as we use the channel statistics through the SNR's CDF for finding $\gamma_0$, we refer to this scheme as the \textit{fading-dependent} approach\footnote{Learning the channel statistics is required \cite{Angjelichinoski.2019}, 
 but done in the long term.}.   
\section{Numerical Results}\label{results}
Herein we present numerical results on the performance of the discussed combining schemes. 
We assume Nakagami-m fading, which allows modeling a wide variety of channels, including Rayleigh ($m=1$) and fully deterministic LOS ($m\rightarrow \infty$) channels. The payload is set to $32$ bytes ($256$ bits), which is  a typical parameter of URLLC use cases with error probability requirements around $10^{-5}$ \cite{3GPP.2017}.
%
\begin{figure}[!t]
	\ \ \ \ \ \includegraphics[width=0.48\textwidth,center]{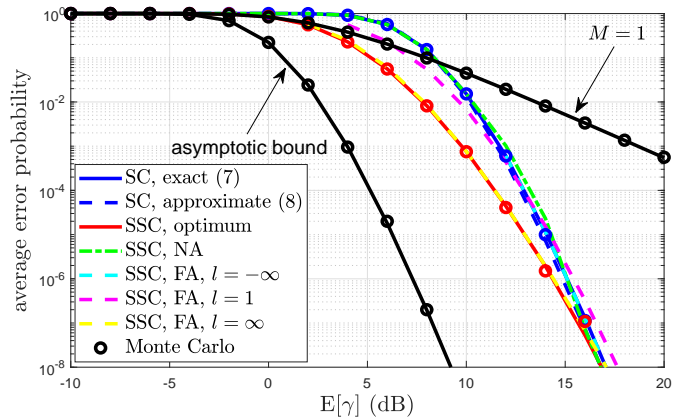}
	\caption{Average error probability as a function of $\mathbb{E}[\gamma]$. We have set $k=32$ bytes, $m=2$, $M=6$ and $u=200$, $p=4$, $q=16$, $d=24$ channel uses.} \label{Fig1}
\end{figure}

Fig.~\ref{Fig1} shows the average error probability of both, SC and SSC, schemes as a function of the average SNR.
%
%
We consider three different methods for computing the SNR threshold for the SSC scheme: \textit{i)} finding the optimum solution of $\mathbf{P}$ in  \eqref{P1} via numerical solvers, which is too costly for a simple IoT device but it is utilized here as benchmark; ii) computing $\gamma_0$ based on the NA approach; and
\textit{ii)} computing $\gamma_0$ based on the FA approach with $l\rightarrow-\infty$ ($\tilde{n}=n_M$), $l=1$ ($\tilde{n}=\tfrac{1}{M+1}\sum_{i=1}^{M}n_i$), and $l\rightarrow\infty$ $\big(\tilde{n}=n_0\big)$. It turns out that the FA with $\tilde{n}=n_0$ provides the closest-to-optimum performance, which is a phenomenon observed also in the remaining figures, while the performance of the NA approach is quite distant from the optimum. 
Notice that a poor choice of $\gamma_0$ could negate the benefits from using multiple antennas such that operating with a unique antenna might become preferable, but SSC can be easily designed to avoid this issue, while reaching near-optimum performance. In fact, it is corroborated that by properly selecting $\gamma_0$,  SSC can easily outperform SC. The performance gap decreases as the average SNR increases, so that for sufficiently large $\mathbb{E}[\gamma]$, e.g. $\mathbb{E}[\gamma]>16$dB, we should set $\gamma_0=\infty$ such that SSC functions as SC.
 We have also shown the asymptotic bound ($F_{\gamma}(2^{k/u}-1)^M$, often utilized for the outage analysis of systems without stringent delay constraints), but it is a loose lower bound for measuring the reliability performance under stringent delay constraints, hence, validating the entire analysis carried out in this work.   

\begin{figure}[!t]
	\ \ \ \ \ \includegraphics[width=0.48\textwidth,center]{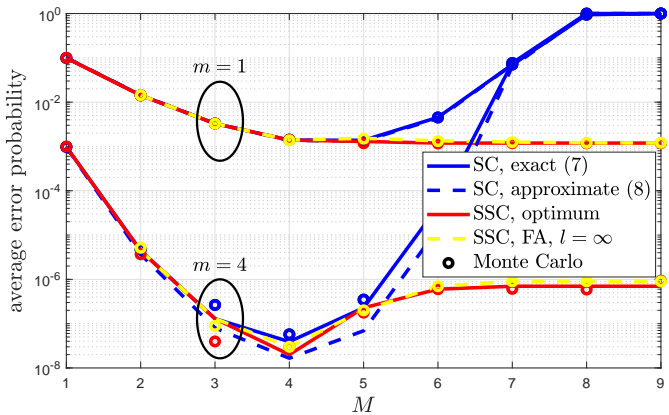}
	\caption{Average error probability as a function of $M$. We have set $k=32$ bytes, $\mathbb{E}[\gamma]=12$dB and $u=200$, $p=4$, $q=16$, $d=24$ channel uses.}
	\label{Fig2}
\end{figure}
Based on above findings, besides the benchmark optimum SSC we just evaluate the FA with $l\rightarrow\infty$ ($n=n_0$)  in the following, while we focus on a setup with $\mathbb{E}[\gamma]=12$dB. Fig.~\ref{Fig2} shows the error performance as a function of the number of antennas for two fading configurations, e.g. $m\in\{1,4\}$. Interestingly, there is an optimum number of antennas under the SC operation. This is because the spatial diversity is limited for small $M$, but for a relatively large $M$ too much time is spent for auxiliary procedures, e.g. $(p+q)M$ channel uses, and consequently the transmit rate increases and the reliability performance is affected as well\footnote{
	For instance,  authors of \cite{Ohmann.2014} reached a similar conclusion regarding an optimum number of links from different transmitters. They determine whether it is more efficient in terms of power to utilize multiple links in parallel (and optimize its number) rather than boosting the power of a stand-alone link. However, notice that such scenario is quite different from ours, in which the cost of using more links is associated to the latency figure. In fact, to the best of our knowledge our observation about an optimum number of receive antennas is new in the literature.}.  The SSC scheme is not severely affected by that situation since by properly setting the SNR threshold it is capable of regulating the number of switches and consequent procedures. Notice that for both SC and SSC the optimum number of antennas is approximately the same. However, while for $m=1$ SSC is capable of providing a near-constant maximum reliability performance for greater $M$, for $m=4$ there is an inflection (which moves to left/right for greater/smaller $\mathbb{E}[\gamma]$) on the average error probability, which causes a slight performance degradation as $M$ increases.
In a nutshell, devices do not necessarily take advantage of a great number of antennas, thus turning off some of them might be required for optimum performance. 

\begin{figure}[!t]
	\ \ \ \ \ \includegraphics[width=0.48\textwidth,center]{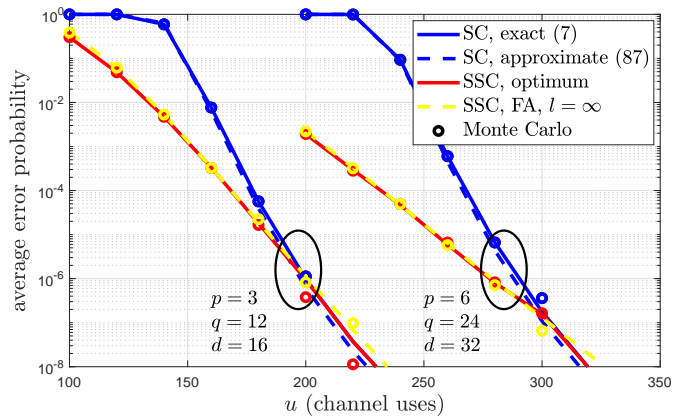}
	\caption{Average error probability as a function of the delay constraint $u$. We have set $k=32$ bytes, $\mathbb{E}[\gamma]=12$dB, $m=2$ and $M=6$.}
	\label{Fig3}
\end{figure}

Meanwhile, the error performance as a function of the delay constraint $u$ is illustrated in Fig.~\ref{Fig3}. Two different setups in terms of $p,q,d$ are evaluated; and as expected, the smaller such values, the better the system reliability since more time can be spent for actual data transmission, which is also favored from relaxing the delay constraint (increasing $u$). Notice that SC performs worse than SSC for relatively small $u$, which suggests that the SC receiver should turn off some of its antennas for a better performance. 

In all previous figures the accuracy of \eqref{Fg} was corroborated, and a slight divergence is observed only when $m$ takes significant values, e.g. $m=4$ in Fig.~\ref{Fig2}, as expected according to our comments before \eqref{e3}. Monte Carlo results validated our analysis in all the cases\footnote{Monte Carlo results are obtained by averaging over $10^7$ samples and consequently the accuracy of the estimated error probability is somewhat affected in the region below $10^{-6}$, as shown in Figs.~\ref{Fig1}-\ref{Fig3}. Notice that error probabilities in the order of $10^{-9}$ may be required in some use cases as commented in Section~\ref{intro}. Such values are reachable via the analyzed diversity combining schemes, specially by using SSC, but under greater average SNR and/or LOS factor. In this work we kept our numerical analysis for error probabilities above $10^{-8}$ since that allowed us to corroborate our results via typical Monte Carlo methods. }.
%
%
\section{Conclusion}\label{conclusions}
We analyzed SC and SSC diversity combining schemes for achieving URLLC when serving a single RF chain IoT device. We demonstrated the necessity of considering the time spent in auxiliary procedures such as synchronization, CSI acquisition, signal processing, feedback transmissions, and switching, due to the harsh time deadline, and show its impact on the reliability performance. 
We demonstrated the superiority of SSC with respect to SC as long the associated SNR threshold is properly selected.
To that end we proposed a fixed SNR threshold which allows reaching near-optimum results by using long-term channel fading statistics. 
We suggest/recommend to turn off some antennas for the greatest reliability performance under ultra-low latency constraints, specially under the SC operation.

%

\section*{Acknowledgment}
This work is partially supported by Academy of Finland 6Genesis Flagship (Grant n.318927, and n.319008, n.307492), FIREMAN (Grant n.326301), and by the Finnish Foundation for Technology Promotion.

\bibliographystyle{IEEEtran}
\bibliography{IEEEabrv,references}
\end{document}